\documentclass[titlepage,11pt]{article}
\pdfoutput=1
\usepackage{amsmath} 
\usepackage{amssymb}
\usepackage{amsthm}
\usepackage{xcolor}
\usepackage[colorlinks=false]{hyperref}
\usepackage{palatino}
\usepackage[numbers,square]{natbib}
\usepackage{fullpage}
\usepackage[multiple]{footmisc}

\title{Report on the NSF Workshop on \\[1ex] \Huge Formal Methods for Security}

\newcommand{\affiliation}[1]{\textit{\small #1}}
\author{
  Stephen Chong\\
  \affiliation{Harvard University}
  \and
  Joshua Guttman\\
  \affiliation{Worcester Polytechnic Institute and MITRE}
  \and
  Anupam Datta\\
  \affiliation{Carnegie Mellon University}
  \and
  Andrew Myers\\
  \affiliation{Cornell University}
  \and
  Benjamin Pierce\\
  \affiliation{University of Pennsylvania}
  \and
  Patrick Schaumont\\
  \affiliation{Virginia Tech}
  \and
  Tim Sherwood\\
  \affiliation{University of California, Santa Barbara}
  \and
  Nickolai Zeldovich\\
  \affiliation{Massachusetts Institute of Technology}
}
\date{August 1, 2016}

\newenvironment{notes}{\itshape\color{blue}}{}

\begin{document}

\maketitle

\section*{Executive Summary}

Cybersecurity is everyone's problem.  The target may be the
electric grid, 
government systems storing sensitive personnel data, intellectual
property in the defense industrial base, or banks and the financial
system.  Adversaries range from small-time criminals to nation states
and other determined opponents who will explore an ingenious range of
attack strategies.  And the damage may be tallied in dollars, in
strategic advantage, or in human lives.  Systematic, secure system
design is urgently needed, and we believe that rigorous
\emph{formal methods} are essential for substantial improvements.
Formal methods enable reasoning from logical or mathematical
specifications of the behaviors of computing devices or processes;
they offer rigorous proofs that all system behaviors meet some
desirable property.  They are crucial for security goals, because they
can show that no attack strategy in a class of strategies will cause a
system to misbehave.  Without requiring piecemeal enumeration, they
rule out a range of attacks.  They offer other benefits too:  Formal
specifications tell an implementer unambiguously what to produce, and
they tell the subsequent user or integrator of a component what to
rely on it to do.  Since many vulnerabilities arise from
misunderstandings and mismatches as components are integrated, the
payoff from rigorous interface specifications is large.  

Adoption of formal methods in various areas (including verification of
hardware and embedded systems, and analysis and testing of software)
has dramatically improved the quality of computer systems. We
anticipate that formal methods can provide similar improvement in the
security of computer systems.

Moreover, formal methods are in a period of rapid development and
significantly broadening practical applications.  While formal methods
have long been associated with cybersecurity applications, new
techniques offer deeper evidence for security goals across a wider
range of components, and for the systems built from them.

Without broad use of formal methods, security will always remain
fragile.  Attackers have a clear advantage in what is currently a
match between the cleverness of the attacker and the vigilance of the
defender.  Formal methods provide guidance for gapless construction,
and for checking that an artifact has no points of entry for the
adversary.  Formal methods always use models, and thus can exclude
only gaps that are expressible in those models. However, each model
has specific, well-defined assumptions, which help focus a security
analyst's attention on whether the actual system satisfies these
properties.

The NSF workshop on Security and Formal Methods, held 19--20 November
2015, brought together developers of formal methods, researchers
exploring how to apply formal methods to various kinds of systems, and people
familiar with the security problem space.  Participants were drawn
from universities, industry research organizations, government, and a
selected pool of scientists from foreign institutions.  We explored
how current research results and strategies can provide improved
secure systems using contemporary formal methods, and how these goals
can shape future refinements to formal methods.

The workshop was organized into four main
areas:  (i) \emph{Hardware architecture}, (ii) \emph{Operating systems},
(iii) \emph{Distributed systems}, and (iv) \emph{Privacy}.  Each area
had an expert area chair (or pair of chairs), who guided
discussion and helped to write a section of the report below.
Participants were assigned to an area for part of the
workshop, with whole group sessions and cross-cutting groups to
consider interactions among abstraction layers.
These discussions led to the following observations, conclusions, and
recommendations:

\begin{enumerate}
  \item Formal methods for security will have an enormous effect in
  the coming years.  Recent advances now enable their use at scales
  that were previously impossible.  The resulting security 
  improvements will spur new investments in formal tools and
  techniques.  
  This interplay will produce a virtuous circle of
  capital investments in the methods and increases in both the quality of
  secure systems and the productivity of security-minded developers.

  \item Formal methods are the \emph{only} reliable way to achieve 
  security and privacy in computer systems.  Formal methods, by
  modeling computer systems and adversaries, can prove that a system
  is immune to entire classes of attacks (provided the assumptions of
  the models are satisfied).  
  By ruling out entire classes of potential attacks, formal methods
  offer an alternative to the ``cat and mouse'' game between
  adversaries and defenders of computer systems.  

  Formal methods can have this effect because they apply a scientific
  method.  They provide scientific foundations in the form of precise
  adversary and system models, and derive cogent conclusions about the
  possible behaviors of the system as the adversary interacts with it.
  This is a central aspect of providing a science of security.

  \item ``Formal methods for security'' should be construed broadly, beyond
  just mechanized logical specifications and proofs. 
  {Formal methods} include approaches to reasoning about
  computational entities in which logical or mathematical descriptions
  of the entities entail reliable conclusions about their behavior.
  Contemporary cryptography relies on formal methods in this broad
  sense, as does synthesis of secure programs and other correct-by-construction mechanisms.
  The broad notion is also particularly relevant for privacy,
  where formal methods naturally extend to rigorous statistical and
  causal analysis methods.

\item Stark challenges remain.
  Computer systems are built in \emph{layers} (e.g.,~hardware, operating
  systems, applications, networking, and distributed algorithms) where each layer is typically built under the assumption that lower layers behave correctly and securely.
  Security may fail at all layers. Frequently, failure is due to 
  mismatches between adjacent
  layers, when behaviors of a lower layer do not satisfy the assumptions
  of a higher layer.  Moreover, different systems (or different
  stakeholders in a system) may seek different \emph{security goals}.
  While traditional goals such as authentication and confidentiality
  are already hard to pin down precisely, privacy goals govern the
  conflict between data subjects who do not want information about
  them disclosed, versus data owners seeking useful or lucrative uses
  for the data.
  
  \item There is no single set of ``right'' security and privacy
  guarantees for computer systems. The desired security and privacy
  guarantees may ultimately depend on specifics of the computer
  application and system deployment.  This heightens the need to
  explore security guarantees rigorously, and particularly privacy
  guarantees.  Privacy should be studied as part of a larger research
  program on personal data protection that encompasses fairness,
  transparency, and accountability.  Hence, formal methods researchers
  should work with researchers in philosophy, law, public policy, and
  related disciplines to forge comprehensive privacy foundations and
  meaningful tools for protecting privacy.

  \item The are many open and compelling research problems, including:
  (1) \emph{Whole-system guarantees}: How to specify and ensure the
  security of a whole system (as opposed to individual components or
  abstraction layers within a system)? How can this be done while still
  enabling modular development and compositional reasoning?
  (2) \emph{Abstractions:} What are the right abstractions to enable
  formal methods for security, including abstractions to present to
  the programmer and abstractions provided by the operating system and
  architecture?
  (3) \emph{(In)Compatibility of Tools, Proofs, and Specifications}:
  To what extent can existing and new tools and techniques be
  standardized to enable compatibility of specifications, proofs, and
  interoperability of tools?
  (4) \emph{Software Development and Formal Methods for Security}: How
  can formal methods for security be supported throughout the
  lifecycle of software and hardware?
  (5) \emph{Transition to Practice}: What is required to enable formal methods
  for security at industrial scales and make them compatible with common
  industry processes?

\item Challenge problems have the potential to ignite the imagination
  and enthusiasm of the community and to 
  stimulate research that pushes the boundary of what is possible
  using formal methods to secure computer systems. We propose
  several challenge problems, including the following:
  \begin{itemize}
  \item \emph{Develop a formally verified crypto-currency wallet.}
    
  \item \emph{Develop an end-to-end secure messaging system on a
    peer-to-peer overlay.}
    
  \item \emph{Develop privacy-preserving tools for scientific
    discovery (data exploration and analysis) 
    by medical researchers, social scientists, and other academics
    working in data-intensive fields for
    daily work.}    
    
  \item \emph{Verify functional correctness of a POSIX-like operating
    system.}
  \item \emph{Use the results to design a post-POSIX operating system
    offering assured security services.}
  \end{itemize}

\item Security and formal methods are both relevant to a broad
  cross-section of the Computer Science curriculum. In undergraduate
  education, security problems should be discussed in a variety of
  courses in which they naturally arise.  Rigorous techniques should
  be introduced relatively early in the curriculum, and connected with
  numerous activities which repay their use.  Graduate education can
  follow suit at a more sophisticated level.

  \item Usable tools and infrastructure are critical to formal methods
  for security.  The community should encourage their development,
  refinement, and shared use.  Possible ways to do so include the
  active encouragement by conferences and journals of the submission
  or evaluation of artifacts for formal methods for security, and the
  establishment of repositories of formal artifacts and
  security-relevant benchmarks and test suites.

\item Clean slate redesigns can liberate innovative, high-quality
  work, but most systems will use much existing infrastructure. A
  balance of both types of work is needed, to provide formal methods a
  clean shot at improving security, as well as a path to broad impact
  by local improvements in existing components.

\end{enumerate}
Thus, we recommend both foundational scientific work and more applied
engineering as foci for improving cybersecurity via formal methods.

\clearpage

\section{Introduction}

\begin{notes}
\end{notes}

Our society depends on an enormous infrastructure of networked
computing systems. These systems have never been well-secured, and as
the payoffs for successful attacks rise, the number and severity of
attacks increase. 
Indeed, in 2015 alone there were tens of thousands of successful attacks
on many parts of US society~\cite{VerizonDBIR2015} including 
the healthcare industry~\cite{breach-carefirst,breach-premera,breach-anthem},
educational institutions~\cite{breach-harvard,breach-penn},
the finance industry~\cite{breach-morgan-stanley},
government and military agencies~\cite{breach-opm,breach-national-guard},
and even computer security specialists~\citep{breach-lastpass,breach-kaspersky}.
The cost of attacks on these systems is estimated to be hundreds of
billions of
dollars~\citep{llyods-ceo,grant-thornton},
not including loss of privacy and damage to national security.

\medskip

Formal methods are approaches to reasoning about computational
entities whereby logical or mathematical descriptions of those
entities enable drawing reliable conclusions about their behavior.
Formal methods enable modeling, verifying, and
synthesizing computer systems.
Formal methods can be usefully applied with varying degrees of
rigor.

Their use to ensure security has been recommended since the 1970s. The
Anderson report~\citep{Anderson1974}, Bell and LaPadula's early work
on operating system security
modeling~\cite{LaPadulaBell73,BellLaPadula73}, and Needham and
Schroeder's 1978 paper on authentication
protocols~\citep{NeedhamS1978} all stressed the importance of rigorous
analysis of detailed models of secure systems.

By explicitly modeling the computer system and the abilities of
adversaries, formal methods can prove that the computer system is secure against \emph{all} possible
attacks (up to modeling
assumptions). This provides high assurance of system security, even against
as-yet-unknown attacks. Indeed, formal methods are the only
currently-known approaches that could provide strong end-to-end
security guarantees: security guarantees throughout the execution of a
system and across abstraction layers.

\medskip
Formal methods at assorted levels of abstraction have had significant
success in securing computer systems. Tools such as
SVA~\cite{criswell:sva}, KCoFI~\cite{criswell:kcofi},
CPI~\cite{kuznetsov:cpi}, and Verve~\cite{yang:verve} have
demonstrated the practicality of formal methods to build systems with
strong security properties such as control-flow integrity, memory
safety, and type safety.  Such properties can be used to provide
comprehensive application security, as in the Ironclad
project~\cite{hawblitzel:ironclad}. There have been significant
advances in proving functional correctness for an OS kernel (including
seL4~\cite{klein:sel4-tocs}, CertiKOS~\cite{gu:certikos-dscal},
ExpressOS~\cite{mai:expressos}, and MinVisor~\cite{dahlin:minvisor})
and of internal kernel components (such as
Rocksalt~\cite{morrisett:rocksalt}, Jitk~\cite{wang:jitk},
FSCQ~\cite{chen:fscq}, and XMHF~\cite{vasudevan:xmhf}).  
Tools such as GLIFT~\cite{TiwariWMMCS2009},
Caisson~\cite{Caisson2011}, Sapper~\cite{Sapper2014},
SecVerilog~\cite{ZhangWSM2015}, and SC-Sniffer~\cite{scsniffer} use
formal methods to ensure strong information security properties of
hardware architecture.
Advances in software-defined networking enable the
synthesis~\cite{McClurgHCF2015} of consistent network configuration
updates~\cite{ReitblattFRSW2012}, which can ensure, for example, that
firewall invariants are never violated, and thus insecure
packet-flows are impossible.
Formal methods also provide secure-by-construction methods for
building systems, including program synthesis (e.g.,~\cite{zznm01,GanapathyJJ2005,HarrisJR2012}).

Contemporary cryptography is an example of a flourishing interplay
between rigorous mathematical methods, a clear model of the adversary,
and strong practical motivations.  Tools based on formal methods can
reason about the correctness of cryptographic algorithms (e.g.,
Cryptol~\cite{Cryptol},
CryptoVerif~\cite{blanchet2008computationally},
EasyCrypt~\cite{barthe2011computer,almeida2013certified},
RF*~\cite{barthe2014probabilistic},
CertiCrypt~\cite{Barthe:2009:POPL}) and produce code or hardware that
provably implements cryptographic algorithms correctly.  Formal
methods were key to discovering the FREAK SSL/TLS vulnerability in
2015~\cite{BeurdoucheEtAl15,beurdouche2015flextls}, which affected
roughly a third of all deployed SSL/TLS servers.

Formal methods have also had recent success in specifying and
enforcing privacy guarantees. Differential
privacy~\cite{dwork-sensitivity06, dwork-privacy06,
  dwork2014algorithmic} provides a strong compositional formal notion
of privacy, and several tools and systems provide enforcement and
verification of differential privacy (e.g., Pinq~\cite{pinq,
  proserpio2012workflow, Ebadi:2015:DPG:2676726.2677005},
Airavat~\cite{roy-2010-airavat}, DJoin~\cite{narayan-2012-osdi},
Fuzz~\cite{fuzz, haeberlen-2011-usenixsec},
DFuzz~\cite{gaboardi2013linear}, VFuzz~\cite{narayan2015verifiable},
GUPT~\cite{mohan2012gupt}, CertiPriv
~\cite{Barthe:2013:PRR:2542180.2492061}). At the interface between
technology and policy, significant progress has been made on formal
specification and enforcement of privacy laws and enterprise policies
\cite{BarthDMN06,DeYoungGJKD10,GargJD11,ChowdhuryJGD14,BasinKMZ15}.
Several industrially-deployed privacy-protection systems are either
directly supported by formal methods or inspired by more formal work.
For example, Microsoft's Bing search engine uses a domain-specific
language, Legalease, to specify privacy policies and a tool, Grok, to
track how user-data flows among programs and check privacy policy
compliance on millions of lines of code written by several thousand
developers.

\medskip

However, significant challenges must be overcome to fully realize the
potential benefits of formal methods for security.  These challenges
concern the scale of the formal methods needed, their integration
across the layers of abstraction of real systems, and their adaptation
to the environments and security goals of systems.

Thus, the National Science Foundation sponsored a workshop on the topic,
to identify existing successes and
opportunities for applying formal methods to security problems, and
raise awareness of these opportunities in relevant communities in
academia, industry, and government research labs. The
workshop was held at the University of Maryland, College Park,
November 19--20, 2015. The workshop had 37 attendees from academia and
industry, and an additional 7 attendees from government
agencies. 

Through a series of discussions and presentations, workshop attendees
identified many exciting open research problems and opportunities and
made recommendations that aim to raise awareness
and encourage useful research and development. 

This report focuses on the motivations for incorporating formal
methods into cybersecurity activities; the opportunities for doing so
and the obstacles; and a variety of challenges and activities that
will enrich the state of practice and of scientific knowledge in key
ways.  Thus, this report is narrower than the recent, well-constructed
\emph{Federal Cybersecurity Research and Development Strategic
  Plan}~\cite{fedcyber16}, which we recommend to readers who may
desire a broader view of the cybersecurity challenges and of research
and development strategy.  

\bigskip

Open research problems identified by the workshop include the following.
\begin{enumerate}
\item \textbf{Whole-system security.} How do we specify and ensure the
  security of a whole system, while still enabling
  modular/compositional reasoning and development? For example: How do
  we specify security guarantees of components to facilitate reasoning
  about security when we combine components? How do we specify the
  assumptions and guarantees of abstraction layers to facilitate
  reasoning about security across abstraction layers?

\item \textbf{Abstractions.} What are the right abstractions to enable
  formal methods for security? These include 
  abstractions to present to the programmer and abstractions provided
  by the operating system and architecture.  Particularly with respect
  to hardware architecture, they also include useful sets of formally
  defined, composable, verifiable, and high performance security
  primitives. Exploration and validation of abstractions is urgent:
  they are central to developing secure computer systems and
  systematically applying formal methods to these systems.

\item \textbf{(In)Compatability of Tools, Proofs, and Specifications.}
  To what extent can existing and new tools and techniques be
  standardized to enable compatibility of specifications, proofs, and
  interoperability of tools?

\item \textbf{Software Development and Formal Methods for Security.}
  How can formal methods for security be supported throughout the
  lifecycle of software and hardware, including supporting security
  goals and mechanisms in the design process, reducing the effort
  required to construct specifications and prove that specifications
  are met, and enabling the continued use of formal methods as a
  system evolves after initial deployment?
  
\item \textbf{Transition to Practice.} What is required to enable formal
  methods for security at industrial scales and make them compatible with common
  industry processes?
  
\item \textbf{Mapping the Space of Privacy.} What is the conceptual
  space of privacy requirements, and is there a computational
  formalization of these requirements (analogous to defining
  complexity classes and the complexity class hierarchy)? How do we
  develop a common framework that accounts for different
  privacy-relevant guarantees?

\end{enumerate}

This report makes the following observations and recommendations.

\begin{enumerate}
\item \textbf{Challenge problems} have the potential to ignite the
  imagination and enthusiasm of the community and to stimulate
  research that pushes the boundary of what is possible using formal
  methods to secure computer systems. We propose the following
  challenge problems.
  \begin{itemize}
  \item \emph{Develop a formally verified crypto-currency wallet.} This challenge emphasizes providing whole-system security for end-user software comprising hardware, an operating system, and application code. The financial relevance of the software provides clear motivation for strong security guarantees, including the characterization and enforcement of privacy and accountability. It may be reasonable to have dedicated or specialized hardware. The security of the crypto-currency wallet likely relies both on the system itself and on properties of the cryptographic protocols.
    
  \item \emph{Develop an end-to-end secure messaging system on a peer-to-peer overlay.} This challenge emphasizes whole-system security for a distributed application. 
  \item \emph{Develop privacy-preserving tools for  scientific discovery (data exploration and analysis) that can be used by medical researchers, social scientists, and other academics working in data-intensive fields to carry out their daily work.} This challenge seeks to connect strong formal notions of privacy with research on real data sets  in social and life sciences.
    
  \item \emph{Verify functional correctness of a POSIX-like operating
    system.} This challenge will push forward the scale of formal
  methods for software verification and help identify suitable
  abstractions that the operating system requires of hardware and that
  the operating system can present to the application to enable the
  use of formal methods for security.
  \item \emph{Use the results to design a post-POSIX operating system
    offering assured security services.}  This challenge will require
  identifying the security goals that a wide range of applications
  achieve, designing an OS interface providing services that allow
  them to achieve their goals, and ensuring that the implementation
  delivers these services.

  \end{itemize}

  \item \textbf{Outreach and Education.}  We need to advocate for the
  advantages of formal approaches, and specifically for formal
  approaches to security, as well as to inculcate a grasp of them
  among newly trained professionals.  
  \begin{itemize}
    \item Outreach advocacy to various communities, including security
    researchers, systems researchers, and hardware designers is
    needed. 

    ``Formal methods'' for security can and should be interpreted more
    broadly than just mechanizable logical specifications. Rigorous
    mathematical or logical methods to reason about the behavior of
    computational entities can help document goals and provide a basis
    for understanding and discussing privacy and security
    requirements. This is perhaps particularly relevant with respect
    to privacy, where formal methods naturally extend to rigorous
    statistical and causal analysis methods, and privacy has been
    extensively studied in diverse disciplines.

    \item Security and formal methods are both relevant to a broad
    cross-section of the Computer Science undergraduate curriculum. As
    such, we recommend incorporation into existing courses of both (a)
    security concepts and techniques and (b) formal methods and
    tools. Although the Computer Science Curricula 2013 \cite{CS2013}
    proposes this with respect to security, it does not do so for
    formal methods. Incorporation of formal methods into existing
    courses (as opposed to the development of new courses focused on
    formal methods) is also the recommendation of the 2012 NSF
    Workshop on Formal Methods~\cite{formalmethodsreport}.

  \end{itemize}

  \item \textbf{Development of Tools and Infrastructure.} The
  availability of tools and infrastructure will be critical to the
  success of applying formal methods to improve security.  We
  recommend developing and sharing tools and other infrastructure to
  enable the application of formal methods.
  
  Based on previous successes of the application of formal methods for
  software correctness, one possible way to achieve this is for
  conferences and journals to encourage the submission or evaluation
  of artifacts, which encourages development and reuse of tools and
  infrastructure. Another possibility is the establishment of
  repositories of formal artifacts and security-relevant benchmarks
  and test suites, to encourage the availability of tools and shared
  infrastructure.

  Much work on tools and infrastructure can pursue an integrated
  transition to practice, which some funding mechanisms can support.
  Additional work to polish tools, infrastructure, and substantial
  worked examples can make them accessible to a broader community
  including systems developers.  Thus, research can lead much more
  quickly to social benefits.

  We note that there is value in both clean-slate redesign and
  incremental improvement to existing infrastructure.  Clean-slate
  redesigns can liberate innovative, high-quality work, but
  improvement to existing infrastructure can have a more immediate
  impact on existing systems.  A balance of both types of work is
  needed.

\end{enumerate}

Section~\ref{sec:goals} describes the goals of the workshop.
The workshop was structured into four main
areas:
Hardware architecture (Section~\ref{sec:hardware}),
Operating systems (Section~\ref{sec:os}),
Distributed systems (Section~\ref{sec:distributed}), and
Privacy  (Section~\ref{sec:priv}).
In addition, discussions were organized around cross-cutting concerns, including
whole-system security guarantees~(Section~\ref{ss:whole-system}), education and outreach~(Section~\ref{ss:education}),
and tools and infrastructure~(Section~\ref{ss:tools}).

\section{Workshop Goals}
\label{sec:goals} 

We had three main goals at this workshop.  First, we wanted to
document some of the central security problems, stretching across a
number of layers from hardware and operating systems through
distributed systems to the more human-oriented questions of privacy.

Second, we wanted to appraise the relevance of today's robustly
developed formal methods.  These are able to handle great complexity
now, particularly when the models are at a fairly uniform level of
concreteness.  Challenging aspects of security are that its concerns
may be logically complex (such as non-interference); its goals may be
hard to formulate (as are many privacy concerns); and systems
may be susceptible to attack at many different levels
of abstraction.

Third, we wanted to identify areas where formal methods are most
likely to make a contribution to security.  These have to be areas
with a history of important security weaknesses; they must be complex
enough to be hard to get right by ordinary careful work; but
convincing models of the crucial security considerations must be
within reach.

Above all, we wanted to provide a forum for interaction among the
extremely varied and strong participants.  Outcomes of real value
include stimulating new collaborations, a new appraisal of the most
pressing problems, a new respect for the available techniques, among
those present.

We proposed five main questions to structure the discussions:  
\begin{enumerate}
  \item What is the evidence that formal methods can make a
  substantial difference to the real practice of security? 
  \item What are the obstacles that could prevent formal methods from
  achieving substantial benefits? 
  \item What are the most promising applications areas and security
  goals?
  \item Why now:  What changes suggest that now is a high payoff time
  for interactions between security and formal methods?
  \item What to do next? (Recommendations/ideas/challenges)  
\end{enumerate}
These questions were applicable across the four areas into which we
subdivided the workshop:
\begin{description}
  \item[Hardware Architecture] led by Tim Sherwood and Patrick
  Schaumont; 
  \item[Operating Systems] led by Nickolai Zeldovich; 
  \item[Distributed Systems] led by Andrew Myers; 
  \item[Privacy] led by Anupam Datta and Benjamin Pierce.  
\end{description}

The area chairs guided discussion and helped to write sections of the
report below.  Participants were assigned to an area for part of the
workshop. In addition, there were several group discussion sessions,
and participants self-organized into areas of cross-cutting concerns,
including whole-system security guarantees, education and outreach,
and tools and infrastructure.

 \section{Area: Hardware Architecture}\label{sec:hardware}

\subsection{Brief overview of the problem area}

Underlying every computing system, from the smallest embedded sensor to the
largest ware\-house-scale distributed system, ultimately is some form of computing
hardware.  All software abstractions---from the application logic, to language
run-time, operating system, and virtual machine---in the end perform their
function through a set of low-level commands to a physical device.  This fact
provides both significant challenges and opportunities to system security.  On
one hand, the hardware sub-systems implement the lowest level of computing
abstraction and cannot be undercut by software implementation artifacts.  By the
nature of sitting at this lowest level, hardware mechanisms have the opportunity
to provide a formally sound foundation on which to build rich and layered
approaches to software security.  On the other hand, hardware is physical and
attackers are not constrained by the formal model that is used to develop or
verify the hardware.  Security failures in hardware may not be easily ``patched''
and may provide complete access to the entire system state.

Despite these significant challenges and opportunities, most hardware designers
today have limited opportunity to learn about system security issues, let alone
having access to formal tools and techniques to help them in their efforts.
Security analysis does not fit cleanly into the existing process for functional
verification and, in some cases, may even negatively impact designers' efforts to
meet performance goals.  Significant research is needed to help bring the power
of formal analysis to bear on the myriad problems of hardware security.

\subsection{Central security goals to achieve}

The physical nature of hardware means that many different classes of
attack are possible.  Each operation pulls a measurable amount of current from the
power supply, each wire toggled emits observable electromagnetic radiation, each
high-energy particle strike opens the possibility for critical bits to be
flipped, and each chip that falls into an adversary's hands is an opportunity to
reverse engineer an entire design.  In some cases one might not even trust the
manufacturing pipeline in its entirety. These classes of physical attack are
``model breaking'' for the vast majority of formal approaches today.  These, and
other, physical attacks must be placed on a more formal foundation and they need
to be considered both independently and in conjunction.

Hardware designs today are often developed against an informal
specification, e.g., an English language document describing the
intent of the design, rather than a mathematical definition of the
operation of the design.  Researchers have increasingly taken to
attempting to ``formalize'' these informal specifications, which they
can then test against the observed behavior on a set of designs.  We
do not need to ``discover'' a formal foundation for hardware.  What we
need are methods to create, analyze, and execute formal hardware
security specifications, and prove equivalences between them,
starting from the earliest points in the design process.

Finally, new security mechanisms are needed to ease the creation and verification
of higher-level system security properties.  Certain properties, such as true
randomness, are available only at the hardware level, while other properties
(such as performance of cryptographic components, isolation, and determinism) may be significantly
improved through additional functionality at the hardware level.  A set of
formally defined, composable, verifiable, and high performance security
primitives has the promise to transform the state of hardware security, and
with it, the bedrock of software security.

\subsection{Evidence that formal methods can help us achieve them}

Formal techniques are well known in general hardware design, especially at the
lower abstraction levels (e.g., layout versus schematic equivalence checking), as
well as in specialized subtasks of hardware design (e.g., finite state machine
reachability testing). Secure hardware design poses particular design challenges
by itself, and formal techniques can improve the design process in two
areas of secure hardware design: to verify the logical properties of a given
design, and to verify its physical properties.

Capturing secure hardware in a Hardware Description Language is error prone.
Instead, domain-specific languages (e.g., Cryptol~\cite{Cryptol}) support abstraction and
verification of synthesized results. For example, secure hardware, in particular
hardware for cryptographic operations, is often based on specialized arithmetic
derived from finite fields, and typically involves non-standard wordlengths.
Formal tools support the design correctness of these highly specialized operators 
by demonstrating the equivalence between high-level specification and the implementation.

Another important area of success in verification is in information-flow
analysis in hardware circuits (e.g., GLIFT~\cite{TiwariWMMCS2009}, Caisson~\cite{Caisson2011}, Sapper~\cite{Sapper2014}),
where high complexity prevents a
designer from doing manual verification. This kind of analysis leads to
a guarantee with respect to isolation. For example, it enables the integration of
trusted and untrusted logic in the same physical chip package.

Through proper modeling, formal methods can verify the physical effects of
hardware execution, including timing and power consumption. With such models, designers can
reason about side-channel leakage (timing and power), and can verify
countermeasures such as constant-time design (e.g., SecVerilog~\cite{ZhangWSM2015}) and
perfect-masking for power randomization (e.g., SC-Sniffer~\cite{scsniffer}). Architecture models 
further help to
verify the physical effects of software execution, such as cache timing effects
(e.g., CacheAudit~\cite{DoychevFKMR13}).

\subsection{Obstacles to the applicability of formal methods}

Several obstacles hinder adoption of formal methods for security in hardware.
First, although there has been significant uptake of formal methods for
functional verification of hardware, functional verification is
typically  performed only piecemeal, on parts of a design.  Security is
often a holistic property of an entire design and questions about the
scalability of these approaches are always present.  Second, hardware
does not currently
have the same open culture as software.  Commercial grade language
run-times, operating systems, and virtual machines are all openly available,
free to study and run, and they are contributed to by a broad community of
researchers.  In the hardware space, most security-critical hardware is not only
closed-source, but it is often so well guarded that it won't be shared even with
trusted commercial partners.  This significantly impedes our ability to
understand the true needs of security-critical hardware and develop innovative solutions.  Finally, the scope
of attack models for hardware is staggering, and includes side-channels, tampering, hardware
Trojans, fault injections, and software-coordinated attacks.  While
these challenges are significant, they are surmountable and can likely be
overcome through a sustained effort from the community and investment from both
government and industry.

\subsection{Promising areas for upcoming research}

Despite the early successes of formal techniques in hardware design, the scale
and scope of the problem domain still have significant room for improvement.

First, there is a great need for tools that can scale with design size. This
calls for better modeling and especially better abstraction of security issues.
Such models should, ideally, capture risk---the product of loss-probability and
cost---over multiple possible threats. In practice, models should initially
focus on accurately capturing the design cost of security (overhead) against the
likelihood of a successful attack. Similarly, models could capture the design
cost of privacy.

A second major challenge is the automatic analysis for Trojans, either
statically or dynamically. Such analysis does not look for bugs in design
artifacts as specified. Instead, it looks for unspecified design
artifacts.

A third major challenge is the development of scalable, formal models to analyze side-channel
leakage, fault propagation behavior and information flow, in particular over
long and extended schedules.

A fourth major challenge is the extension of formal properties derived from
hardware into software. Indeed, many interesting cases of high-assurance design
do not involve isolated software or hardware, but rather a combination of them.
A closely related challenge is the verification of customized microprocessor
features that enforce security properties such as isolation, memory
confidentiality, guaranteed service, and so on.

Finally, reconfigurable and runtime-adaptable systems will need formal proofs
that can be adapted at runtime. In addition, protocol features such as nonces
and truly random inputs can be verified only at runtime. Formal techniques could
help in both of these cases to reduce the cost of runtime testing.

\subsection{Is this area ripe for a fresh focus?}

While hardware has always played an important role in software security, (e.g.,
through memory management via the TLB, with support for virtualization through
trapping, etc.) the hardware/software interface remained relatively static for
many decades.  However, due to the continued slowing of transistor
power/performance system there is a radical transformation now taking place.
Systems are becoming increasingly parallel, decentralized, heterogeneous, and
rich with custom hardware functionality.  For the first time in many years,
programmers are being asked to understand and explicitly manage the underlying
hardware in a new way.  This shift means that entirely new blocks of the system
(e.g., on-chip networks and transaction memory processing hardware) are asked to
play a significant new role in security. The challenge is that the security
properties of these diverse new architectures are not well understood but the
opportunity is that software developers are more open now than ever before to
changing the fundamental hardware/software contract.

\subsection{Actions that can create momentum}

Formal tools for secure hardware design face the same challenges as other formal tools
for hardware design: they are not well integrated with the common hardware
design flow, especially at the higher abstraction levels. However, the context
of security offers several compelling advantages for the use of formal tools, as
argued above. Thus an important driving force in creating momentum in this area
will be to create opportunities to bring the secure hardware design community
and the formal verification community closer together.

First, there is need for open repositories that describe real design artifacts
and actual, real-life security problems. The secure hardware design community
has been very successful, for example, in stimulating research in side-channel
analysis through open design artifacts, measurements and hardware (e.g.,
DPA-contest~\cite{Clavier2014}, SASEBO-II side-channel analysis board~\cite{sasebogii}). To rally the formal
community into the challenges of secure hardware design, a set of common
benchmarks is needed. Some benchmarks are already available (e.g., Trust-Hub \cite{trusthub}),
but there is need for a structure that enables the formal verification community
to interact with them.

Second, there is need to advocate the advantages of the formal approach in the
hardware design community, and vice versa, to explain the challenges of secure
hardware design to the formal community. This could be done by engaging the
leading researchers of each field to host tutorials or invited talks at the main
conference venues of these communities. A Distinguished Speaker program could
help to support the travel costs and engagement costs for these speakers.

Third, there is a need to engage industry to prioritize the many security
challenges faced in a complex chip design. This can be done by engaging industry
consortia such as SRC, and by involving them in the research program of NSF
(similar to, or as part of, SaTC/STARRS, but with an emphasis on formal tools
for security).

\section{Area: Operating Systems}\label{sec:os}

Operating systems provide many services that applications rely on, such
as a network stack, a file system, process isolation, inter-process
communication, and so on.  
The security of applications depends pervasively on the underlying OS,
making operating systems an appealing target for applying formal
methods.

For instance, if a system executes multiple applications (or virtual
machines) on the same computer, the OS kernel is responsible for ensuring
that a malicious (or compromised) application is unable to tamper with
the execution of other applications.  Even if the computer is used for
running just one application, OS-level process isolation is often used to
isolate less-trustworthy components of a large application to mitigate
the damage from a potential compromise.  Ensuring the correctness of
process isolation in an OS kernel could provide stronger assurance that
compromised components cannot tamper with the rest of the system.

As another example, applications can store user information, such as
passwords, in a file system.  These applications rely on the OS to not
disclose that data, and to not allow an adversary to tamper with the user
passwords (e.g., by changing them to a password that the adversary knows).
Here, the OS might not be directly in charge of enforcing application-level
security, but if the OS functions incorrectly, an adversary can still
subvert the application's security.  Thus, a formal guarantee that the
file system in an OS is working correctly is critical to ensuring that
this application achieves its own security goals.  Applications may
similarly rely on other OS subsystems for their security.

The ultimate goal of applying formal methods to an operating system is
to help application developers build secure applications on top of that
OS\@.  This places a significant emphasis on the interface between the
OS and the application, and on formal specifications of that interface
that would be most helpful for an application developer to prove their
application's security.

One reason why operating systems are an especially appealing target
for applying formal methods is that the same operating system is often
shared across a wide range of applications.  As one example, the Linux
kernel runs on everything from sensors and watches, to mobile phones and
laptops, to high-end servers.  This means that the significant effort
of formally verifying the correctness or security of an operating system
can be amortized across a large number of applications that would benefit
from the verification effort, and thus potentially provide a high payoff.

As we will discuss shortly, there are exciting success stories
in applying formal methods to operating systems.  However, there is
still no formally verified operating system that provides the typical
services applications expect, such as a file system, a network stack, etc.
Building such an operating system, and using it to develop examples of
secure applications on top of it, is an important next step for this area.

\subsection{Benefits of formal verification}

One of the main benefits of applying formal methods to OS kernels is
that it can provide a strong assurance of the kernel's correctness or
security properties.  This can be especially useful to kernel developers
working on tricky, bug-prone code inside an OS kernel, such as ensuring
crash recovery in a file system, dealing with concurrent code (such as
Read-Copy-Update in the Linux kernel), and so on.

However, even if the kernel is bug-free, formal methods can have a range
of other benefits.  First and foremost, a precise specification of the
kernel's behavior can eliminate  disagreement between the application
and kernel developers about what an interface provides; such disagreements
have often led to application bugs in the past~\cite{pillai:appcrash}.
This, in turn, can help application developers build secure applications.

Moreover, formal specifications provide a strong way of documenting the
assumptions that an OS kernel is making (e.g., about how the underlying
hardware is behaving, about how the kernel is configured, or about how
the application is using the OS kernel).  The specification can also
help make explicit what critical invariants must be maintained in order
for a particular property (e.g., process isolation) to be enforced.

Finally, formal methods can help make the OS kernel more evolvable.
Having precise specifications frees kernel developers to pursue more
aggressive refactoring or optimizations, since they can be sure their
new code meets the same exact guarantees as the previous version.
Furthermore, precise specifications can enable developers to
add extensions to existing kernel code, without having to worry about
forgetting some subtle detail or interface.

\subsection{Goals for formal methods and initial successes}

One of the central questions in applying formal methods to operating
systems lies in deciding what specification should be proven about the
operating system.  At a high level, there are a number of different
properties that can be proven, from weaker to stronger:

\begin{itemize}

\item  Absence of certain kinds of bugs, or resistance to certain classes
of attacks.  For instance, an OS developer may want to ensure control flow
integrity, memory safety, or type safety for their OS kernel.  One benefit
of providing these properties is that they correspond to significant
classes of attacks in practice, and thus can eliminate certain avenues of
attack.  Another benefit is that it is often possible to check or enforce
these properties for existing code, allowing for incremental adoption.
For example, SVA~\cite{criswell:sva}, KCoFI~\cite{criswell:kcofi},
and CPI~\cite{kuznetsov:cpi} have shown that it is already practical to
build systems that provide these types of guarantees.  Type- and
memory-safety is also a key building block for proving higher-level
properties; for instance, the Verve kernel~\cite{yang:verve} proved
type- and memory-safety of a simple OS kernel, which was later extended
in the Ironclad project~\cite{hawblitzel:ironclad} to prove comprehensive security for the application.

However, these properties protect only against certain classes of
attacks.  They do not provide any guarantees if the adversary uses a
different sort of attack, for instance, enforcement of a control-flow
integrity property does not provide any guarantees against data-flow
integrity attacks~\cite{ChenXSGI2005}.  

\item  Functional correctness of internal kernel components.
This means that the developers have formalized the specification of
some subsystem in the OS kernel, and proven that the code in that
subsystem meets their specification.  This can be particularly
useful for security-critical subsystems, where formal methods
can ensure the absence of bugs in that specific part of the kernel.
For instance, Rocksalt~\cite{morrisett:rocksalt} proved the security of a
Native-Client-like software fault isolation system; Jitk~\cite{wang:jitk}
proved the security of the Seccomp/BPF bytecode interpreter in the Linux
kernel; FSCQ~\cite{chen:fscq} proved the correctness of a file system;
and XMHF~\cite{vasudevan:xmhf} verified internal invariants for
an x86 hypervisor.
Microsoft's SLAM model checker~\cite{BallLR2011} uses predicate
abstraction to scalably analyze the correctness of Windows device
drivers.

Verification of internal kernel components is an important step towards
applying formal methods to an entire OS, and results from this space will
likely help verify the correctness of the same kinds of components in the
context of an entire operating system.  Verifying individual components
can also be a useful strategy for incremental adoption, especially for
security-critical components of existing systems.

\item  Functional correctness of the entire operating system.  This
means specifying the behavior of the entire OS kernel that's visible
to user-space applications (such as system calls, scheduling, etc.),
and proving that the OS kernel implementation meets that specification.

There has been significant work in proving functional correctness
for an OS kernel, with different kinds of user-level interfaces
and corresponding specifications.  For example, the seL4
microkernel~\cite{klein:sel4-tocs} has a proof that its kernel
implementation meets an abstract model of a microkernel, along with the
capDL language for formally reasoning about the isolation properties
achieved by seL4's capabilities~\cite[\S6.1]{klein:sel4-tocs}.
Several other projects have also proven the correctness of simple
hypervisor-like OS kernels, including CertiKOS~\cite{gu:certikos-dscal},
ExpressOS~\cite{mai:expressos}, and MinVisor~\cite{dahlin:minvisor}.
However, researchers have not yet been able to build a provably correct
OS kernel providing traditional abstractions expected by applications,
such as a network stack, a file system, inter-process communication, etc.

\item  As mentioned earlier, the ultimate goal of formal methods would
be to reason about an entire system, consisting of both the operating
system and the applications running on top of it.  Here, the ultimate
specification is application-dependent, and the OS specification serves
only as a way to help the application developer prove that the application's
own specification is satisfied.  The most prominent result in this space
is the Ironclad project~\cite{hawblitzel:ironclad}, which proved the
correctness (and security) of several applications, including a password
hasher and a differentially private database.  seL4's work on CapDL has
also been used to reason about the isolation of programs running on top
of the seL4 microkernel~\cite[\S6]{klein:sel4-tocs}.  Finally, some work has
been done on proving the security of simple applications on top of existing
operating systems with the help of shims~\cite{ricketts:reflex, jang:quark}.

\end{itemize}

\subsection{Open research questions}

\paragraph{Factoring out security.}

In the context of applying formal methods to operating system security,
one of the biggest questions is, what security primitives should the OS
provide to its applications?

Ideally, the OS would provide applications with mechanisms that factor
out application security from other (non-security) correctness concerns.
This would, in turn, allow application developers to focus their formal
method efforts on the security of their application (where formal methods
may be able to add significant value), and less on the overall correctness
(where applying formal methods may be of less value).

Alternatively, if such mechanisms do not exist, then there is little
difference between full functional correctness and security at the
OS interface.  Thus, applying formal methods to OS security will
mean specifying and proving full functional correctness of the OS
interface, and it will be up to application developers to prove their
application-level security goals, on top of the OS kernel's functional
correctness specification.

Unfortunately, there isn't clear agreement on this question, even
without considering formal verification.  Existing security mechanisms
such as access-control lists (ACLs) are widely deployed, but do not seem to provide much help
in reducing the effort of reasoning about security.  Information flow
control (IFC) and capabilities are two alternative security mechanisms
that have been widely studied, and in principle can help application
developers reduce the code that has to be considered for the purposes
of security properties.  However, it's not yet clear how hard it is to
build large-scale applications using IFC or capabilities in practice,
although some initial evidence suggests some variants of IFC may be
promising~\cite{stefan:cowl, giffin:hails}.

\paragraph{Specifications.}

OS interfaces such as POSIX have traditionally been specified
informally, using English at best.  As a result, the specifications
may not be a good fit for formal methods, which require a precise
description of how an interface operates.  What's the right approach to
formalizing OS interfaces? Should we formalize POSIX despite unclear or
inconsistent handling of corner cases, which might lead to needlessly
complex and hard-to-use specifications? Should we start anew with a
specification geared towards formal verification from the beginning?
Or is there a way of evolving existing interfaces like POSIX to be more
formalization-friendly?

\paragraph{Hardware models.}

OS kernels run on bare hardware, which means that formally reasoning
about an OS kernel requires a formal model of the underlying hardware.
Building such a model is a non-trivial task: processors are highly
complex, especially when considering the privileged instructions
needed by an OS kernel, and the rest of the hardware platform used by
an OS kernel (DRAM controllers, timers, PCI, power management, devices,
etc.) also requires formalization.  One approach taken by prior work is
Ironclad's idiomatic specification~\cite{hawblitzel:ironclad}, which
formalizes just the subset of the instruction set that is actually used.
However, for an OS kernel that can run arbitrary user-space code, can
this approach still work?  And how feasible is idiomatic specification
for the rest of the hardware platform, aside from the CPU?

\paragraph{Concurrency.}

Concurrency is a key concern for OS kernels, which are typically in
charge of running multiple processes on a single computer.  However,
most existing work on OS verification focuses on sequential execution
of OS kernels.\footnote{The one exception is Microsoft's work on using
VCC to verify the Hyper-V hypervisor~\cite{cohen:vcc-local}, although
that project stopped before they were able to finish the verification.}
Moreover, concurrency is fundamentally required to reason about the
execution of multiple processes or applications on top of the same
OS kernel.

However, formal reasoning about concurrency seems to be still in the early
stages; there is still no consensus on the best way to reason
about concurrent programs, or how to verify that concurrent code meets a
specification.  Perhaps the biggest formally verified concurrent program
is a garbage collector~\cite{hawblitzel:civl}.  Addressing these basic
questions is critical to make progress on OS verification.

\paragraph{Systems programming language.}

What language should the verified system be written in?  C is well
understood and low-level, but has complex semantics; the resulting proof
obligations can significantly increase the proof effort.  The C language
is also not well integrated with formal tools such as proof assistants,
making it more difficult to co-develop code, specifications, and proofs.
Functional languages like Haskell or Gallina are better integrated into
formal tools, so that it's easy to change the code, specification, and
proof all in the same file and same development environment.  But can
functional languages provide acceptable performance for an OS kernel?
Newer languages such as Go and Rust provide a potential
alternative, although it's not yet clear whether Go's garbage
collector is compatible with the performance needs of an OS kernel, or
whether Rust's concurrency memory model is a good fit for an OS kernel
that fundamentally operates on shared memory.

\paragraph{Development effort.}

A significant barrier to adoption of formal methods in OS development is
the high development effort.  How can we reduce the effort required to
construct specifications and prove that code meets them?  Perhaps even
more importantly, how can we make sure that future changes to the OS
kernel don't require developers to redo all of the proofs?

\paragraph{Clean-slate or incremental deployment.}

What's the best route to making sure that work on formal methods in
operating systems achieves real-world impact?  Clean-slate approaches
offer appealing simplicity, yet make it difficult to deploy in an existing
system.  One answer may be to verify individual components of existing
systems incrementally, as described earlier; however, to achieve full
functional correctness of an OS in the long term, it is important to
eventually combine these individual components into a comprehensive proof
for the entire OS\@.

\paragraph{Whole-system verification.}

Ultimately, verifying the OS kernel is just a step to proving
strong properties about the entire system, which includes the
OS kernel and the applications running on top of it.  But how
can we verify the entire system, when the applications (and the
kernel) might be written in different languages, with different
styles of theorems and specifications, and different proof tools?

Even within an OS kernel, different methodologies, proof techniques,
or formal tools, might be best suited to different parts of the kernel
code.  How can we effectively combine them?

We explore some of these questions separately in Section~\ref{ss:whole-system}.

\subsection{Possible next steps}

We believe that the next goal in OS verification should be to actually
prove functional correctness of a complete POSIX-like operating system.
Doing this would require addressing many of the research challenges
listed above, such as coming up with a specification for a reasonable
subset of the OS interface, determining the best approach for handling
concurrency, developing a suitable hardware model, choosing a language
for implementing the OS kernel, and actually proving its correctness
against the specification.

One way to drive this work, and in particular, to ensure that it produces
a useful specification, is to focus on an example application that
would stress the need for a usable formalization of the OS interface.
For example, building a verified Dropbox-like file sharing application,
or a provably secure banking app for a cell phone, or provably secure
encrypted off-the-record messaging, would be a good driver for the
underlying OS verification research.

We expect that the above goal entails a significant amount of research.
Consequently, it may be fruitful to focus on smaller first steps towards
that goal, which would also be useful in their own right.  For instance:

\begin{itemize}

\item Provably correct building blocks, such as an append-only log,
a cryptographic key storage module, persistent storage system, a
high-performance VMM, and a library of concurrent data structures.
These building blocks are likely to be useful in building a comprehensively
verified OS; would likely result in important research outcomes; and
would also be useful in existing systems.

\item Provably correct libraries, such as a verified TLS
implementation, or a library for authorization models (e.g., RBAC),
would both address current problems in these security-critical
libraries, as well as be eventually useful in a comprehensively
verified system.

\item Provably correct execution of a simple system under aggressive
threat models, such as an imperfect memory (reflecting row hammer
attacks~\cite{KimDKFLLWLM2014}) would advance research on hardening software against hardware
errors and attacks, and enable security researchers to provide strong
formal guarantees in these challenging environments.

\end{itemize}

\section{Area: Distributed Systems and Networks}\label{sec:distributed}

Security matters at every layer of modern computing systems, but
especially at the level of distributed systems and networks.  Modern
computing systems and modern applications are typically distributed
systems, with data storage and computation happening at different
nodes in the distributed system. On the user side there are a variety
of different devices ranging from desktop computers to smartphones;
but the functionality of these devices are backed by cloud-based
storage and compute nodes. Distribution is not merely used to connect
users to remote resources; within and across enterprises, services are
stitched together across the network to form larger systems.

The security of distributed systems is critically important. There are
many distributed systems whose compromise could lead to loss of life
include power grids, government and military information systems,
medical information systems. And many more systems are economically
critical.

Unfortunately, distributed systems are hard to secure. Arguably they
pose the greatest challenge:

\begin{itemize}
\item 
\textbf{The whole stack.}
Their security rests on the
security and correctness of every layer of software and hardware below
them. Hence, distributed systems are at least as difficult to secure
as application software, networks, operating systems and hardware.

\item
\textbf{Size.}
Distributed systems are large systems, frequently involving
millions of lines of code or more. These systems are too simply large to
directly construct proofs of correctness or security, although
formal proofs can be applied to key components.

\item
\textbf{Complexity.}
Distributed systems tend to contain complex algorithms that are hard
to implement correctly. Their security often rests on the correctness
of complex cryptographic protocols and fault-tolerance protocols that
are challenging to prove correct even in the abstract, and that
furthermore are easy to implement incorrectly.
\item
\textbf{Evolution.}
Distributed systems are moving targets: they evolve over time. They
are often built using existing service components exporting an API that
can be used by components developed later. Security verification
cannot be done once at the beginning of time; systems must be
reverified as they evolve.
\item
\textbf{No central control.}
Since distributed systems often cross organizational boundaries, any
one participant in a distributed system has less control over the
system as a whole. Many activities are happening concurrently on
modern distributed systems, some not under the control of any given
participant.  Other participants may have their own security goals and
some parts of the system are likely to be opaque to any given
participant.  And it is typically more difficult to exclude ``the
adversary'' from a distributed system, because adversaries have
network access.
\end{itemize}

Although distributed system security offers serious challenges,
the security problem becomes easier in this context in some ways. Current
hardware and software architectures tend to provide a degree of
isolation between distributed nodes ``for free.'' The expectations
of security are sometimes lower for distributed systems. Proofs of
functional correctness may be infeasible, but may also not be
necessary, at least in the near term. It would be a step forward
for many systems if even simple security guarantees could be offered.

\subsection{The value of formal methods for distributed systems security}

There is already considerable evidence that formal methods developed
in the academic research community can have an impact on the security
and correctness of fielded distributed systems:

\begin{itemize}
\item
Engineers at Amazon Web Services have used formal methods~\citep{Newcombe:2015:AWS:2749359.2699417} including
formal verification and model checking to verify the correctness of
their widely used Simple Storage System (S3). They used formal
specifications written in the TLA+ specification language.

\item During the past year, the so-called FREAK vulnerability was
discovered in roughly a third of all deployed SSL/TLS
servers~\cite{BeurdoucheEtAl15}. This rather shocking discovery
depended on the use of formal methods. Researchers at INRIA, MSR, and
IMDEA developed a formally verified TLS implementation that was then
used as a reference implementation against which to systematically
test existing TLS implementations when subjected to deviant message
sequences.  Formal methods contributed to replacing the OpenSSL state
machine with a corrected version.  In fact, formal methods have been
crucial to the development of many secure cryptographic protocols in
use today.  Many protocol suites have been scrutinized using formal
methods, e.g.,~the standardized ISO/IEC authentication
protocols~\cite{BasinCM13}.

\item 
Facebook Infer is a static analyzer developed at Facebook, used by
Facebook engineers to identify null pointer access and resource leaks
in Java programs. Facebook Infer builds on the key technology of
separation logic, which enables precise but scalable reasoning about
program code that performs complex heap manipulation. This system
has also recently been released as open source.

\end{itemize}

\subsection{Challenges for formal methods for distributed system security}

Distributed systems are typically too large and complex to perform
verification after construction or to treat verification as a
monolithic, one-time process. The scale of these systems demands the
development of methods for security assurance that are more modular,
compositional, and incremental. 

Modularity is needed so that formal methods can be applied to
individual system components rather than requiring that the
verifier confront the entire complexity of the system at once.
It must be possible to prove that individual components provide
the properties required of them by the rest of the system, while
treating the rest of the system in an abstract way.

Modularity also demands compositionality: if separately verified
components are combined to form a larger system, the desired security
properties of the larger system should follow from the formally
verified properties of the individual component modules, rather than
requiring that modules be verified again for their new context.

Since the components of distributed systems often lie in different
administrative domains---they are \emph{federated systems}---the
implementations of some components may not be available to be studied
formally.  Participants may only know what security guarantees are
offered (that is, promised) by components not under their control.
Compositional reasoning is therefore crucial to federated systems.

Existing methods for modular, compositional reasoning about
distributed system security are far from satisfactory.  Further,
distributed systems are built using previously implemented services
(possibly in different trust domains) that communicate over a
networked API. An additional challenge is posed because these systems
are constantly evolving. Therefore, it is desirable to have methods for
incrementally verifying distributed systems, so that the work of
verifying security is proportional to the degree of change in the
system being verified, rather than to the total size of the system.

\paragraph{Toward security by construction}

What the past 10 years of increasing success with applying formal
methods to building secure and reliable systems has shown is that
formal methods are most effective when they are part of the design
process---when formal methods are used to capture the evidence and
reasoning that the programmer constructs as part of the development
process.  If software is constructed through conventional means with
this evidence effectively erased after construction, proving important
properties of the software becomes far more difficult.  Formal methods
such as program analysis and model checking can still be applied, but
these methods are currently not modular or scalable.

The existing abstractions and APIs for distributed programming are
also too low-level to support formal verification well. Programming at
the level of, say, TCP sockets not only offers the programmer many
opportunities to make security-critical mistakes, but also obscures
the higher-level security issues. There is simply too large a semantic
gap between these low-level abstractions and the security and privacy
goals of distributed systems. Formal methods will likely become easier
to apply to application code if distributed systems are built using
higher-level abstractions. Higher-level abstractions separate the
problem of security verification into two problems: first, verifying
the implementation of the higher-level abstractions, and second,
verifying the application code that is built using them.

A serious impediment to this goal of higher-level abstractions, and to
modular and compositional verification of distributed systems security
generally, is that the security requirements of distributed systems
are hard to specify and hard to formalize. More research is needed on
ways to capture these requirements in a way that can be presented to
both developers and verifiers. Various promising methods have been
developed for describing at least some aspects of the security of
complex distributed systems. Examples of compositional specification
methods include information flow control, session types, and
separation logic. However, these languages and logics are not able
to capture the full range of security requirements.

\subsection{Goals}

In the longer run it is critical to make progress toward verifiably
secure distributed systems, because too much is at stake. There are
large challenges that likely must be overcome to build verified secure
distributed systems.

Clearly there is need for secure compositional distributed and
cryptographic protocols that deal with heterogeneous trust,
consistency issues, and side channels. More broadly, the community is
still seeking appropriate abstractions to enable the design and
efficient implementation of secure distributed systems. There are many
requirements on such abstractions. The abstractions should expose
security properties (including confidentiality, integrity, and
availability properties) in a form understandable to ``normal'' (i.e.,
non-security-specialist) programmers. We need high-level abstractions
that are suitable for programming ``the Internet Computer,'' that is,
to easily map code and data to distributed systems with heterogeneous
trust. However, the abstractions should not create side channels,
vulnerabilities, or unacceptable performance issues. Ideally,
abstractions for building secure distributed systems should cleanly
interface with ``lower-level'' abstractions (i.e., OS-level
mechanisms) to provide a separation of concerns with respect to
distributed system security versus single-machine security.
Traditional adversary models for distributed systems may need to be
extended to incorporate economic and game-theoretic adversary models,
and, for example, ensure distributed protocols and systems are
incentive-compatible with expected adversaries.

\bigskip

In the shorter term, there are steps that the research community can take
to help build the foundations to solve the larger problem of verified secure
distributed systems.
Higher-level abstractions for constructing and verifying distributed
systems will rely on core building blocks that have been carefully
verified. Unfortunately these core components are largely absent at
present. In many cases, the incentives to both academics and industry
to create verified implementations of these components are currently
too weak. Some examples of needed key building blocks include both
security mechanisms for distributed systems and implementations of
other distributed algorithms:

\paragraph{Secure distributed security mechanisms}
\begin{itemize}
\item Secure authenticated channels are a core abstraction.
      TLS is an attempt to provide such channels, but verified
      implementations are needed.
\item Verified implementations of cryptographic libraries that
      can be used in a composable fashion.
\item At the root of security mechanisms for authorization and audit
      is unspoofable identity. Trustworthy identity management services
      would provide a solid foundation for a wide range of other
      security mechanisms.
\item Multiuser systems must decide whether to authorize requests. In
      the distributed setting, secure authorization becomes more
      difficult: authorization itself may be a distributed computation
      that may be subverted by adversaries or may leak information to
      them. Abstractions and implementations are needed for
      distributed authorization.
\item A functionality growing in importance is the ability to run
      code in trustworthy fashion on untrusted compute nodes.  Support from
      hardware (e.g., SGX) or cryptography (e.g.,  homomorphic encryption) is
      required. Both are areas of active research, but verified
      implementations are needed.
\end{itemize}

\paragraph{Secure distributed algorithms}

\begin{itemize}
\item Consensus is a key distributed algorithm that lies at the
      heart of distributed transaction processing systems and other
      distributed algorithms. For example, current cryptocurrency
      mechanisms are essentially a very inefficient consensus
      algorithm.  An efficient, secure implementation of consensus is
      needed with clearly defined, verified security properties.
\item Many distributed algorithms depend on measuring time, but the
      measurement of time is itself a distributed protocol that could
      be subverted by adversaries. NTP, the standard time protocol, is
      based on strong trust assumptions.
\item When data is stored in faraway data centers, access latency
      interferes with many applications. Replicating the data at
      multiple locations is crucial so that users are typically close
      enough to at least one replica. However, programming with
      replicated data is quite challenging because replicas can become
      inconsistent with each other. Further, the more replicas there
      are, the more likely it is that one is compromised.  It would be
      very valuable to have verified implementations of replicated
      storage abstractions that offer guarantees regarding data
      integrity and the availability and latency of data access.
\item Beyond simple storage abstractions, applications need
      higher-level functionality for accessing remote storage,
      such as atomic transactions and queries.
\end{itemize}

\subsection{A multicommunity effort}

Distributed systems security is a big problem that involves expertise
from multiple research areas: systems researchers (distributed
systems, networking, operating systems, databases), formal methods and
programming languages researchers, and cryptographers. It seems hard
to make real progress on this important problem without bridging the
gaps between these research communities, and may require explicit
action to build community around larger efforts.

\section{Area: Privacy}\label{sec:priv}

In this section, we elaborate on the scope of the research area on formal methods for privacy. We interpret both ``privacy'' and ``formal methods'' in a broader sense than their typical interpretation in the computer science community.  

\subsection{Defining Privacy and Formal Methods}

Privacy has become a significant concern in modern society, because,
increasingly, a wide range of organizations collect, use, and share
personal information about individuals. The emergence of sophisticated
statistical methods for big data analytics, including machine learning
methods, has further exacerbated the problem. Indeed, the very
question of what ``privacy'' means has been extensively studied---and
remains highly contentious---in many disciplines ranging from
philosophy to law to public
policy~\cite{Brandeis,Westin,Nissenbaum,Solove}. Recognizing this
plurality of ideas, we suggest a broadening of work in computer
science on this topic.

A starting point for work in privacy is ensuring the lack of
``inappropriate'' flows of personal data. The determination of which flows
are inappropriate is a difficult normative question. Some have argued for
``privacy as control'' where data subjects decide for themselves how their
data flows~\cite{Westin}. Others have argued for ``privacy in context''
where entrenched norms of a context determine whether a flow is
appropriate~\cite{Nissenbaum}. These are but two examples from an extensive
body of work. Our point in mentioning them is to highlight the fact that
depending on the conception of privacy that is being formalized, different
types of formal methods may be appropriate. At the same time, this body of
work typically views data types as atomic.
    Advances in machine learning and other
statistical methods have been the basis for numerous attacks demonstrating
that seemingly innocuous data types (e.g., an individual's movie ratings, or
social network) can reveal information about other data types (e.g., their
identity or sexual orientation). Thus, nuanced models of information,
statistical inference methods, and related ideas from computer science also
inform the foundations of privacy.  We recommend that computer scientists,
in general, and formal methods researchers, in particular, work with
researchers in philosophy, law, public policy and related disciplines to
forge comprehensive privacy foundations and meaningful tools for protecting
privacy.

A second form of broadening that we suggest is to study privacy as part of a larger research program on personal data protection that encompasses fairness, transparency, and accountability. This viewpoint is consistent with conceptions embodied in the Fair Information Practices Principles (FIPPs)~\cite{oecd1980} and in recent reports from the White House~\cite{whitehouse2014bigdata}.    

We also suggest that the term ``formal methods" when applied to privacy be interpreted more broadly than its typical use. 
In particular, formal methods in specifications should include not just
readily mechanizable logical specifications but 
rigorous methods more broadly, e.g., ones couched in the ordinary mathematical language of statistics.  
Such precise specifications can help documentation of goals, and as a basis for understanding and discussion of privacy 
requirements. Indeed, in a later section we formulate the goal of developing a map of the privacy space as a grand challenge. 
With a similar philosophy, we suggest that the scope of formal methods for enforcement should include a broad class of rigorous methods. 
Examples of such methods are conventional formal methods such as language-based methods, theorem-proving, model checking, 
run-time verification, and unconventional formal methods, such as forms of experimentation and testing of personal information 
processing systems that draw on statistical and causal analysis methods~\cite{datta2015opacity,datta2016transparency}. 

In summary:
\begin{enumerate}
\item Computer scientists in general, and formal methods researchers in
particular, should work with researchers in philosophy, law, public policy
and related disciplines to forge comprehensive privacy foundations and
meaningful tools for protecting privacy.
\item Privacy should be studied as part of a larger research program on
personal data protection that encompasses fairness, transparency, and
accountability.
\item The term ``formal methods" when applied to privacy should be
interpreted more broadly than its typical use to encompass a range of
specification and enforcement methods---in particular, rigorous statistical
and causal analysis methods.
\end{enumerate}

\subsection{Early Successes}

Since the formal study of privacy is a relatively young area, one might expect
success stories to come mainly from academia, and indeed there are many of
these.  We mention some notable ones that reflect successful basic research,
plus some success stories that are indicative of transitions from basic
research to industry practice.

\paragraph*{From philosophy and law to computer science.}

We summarize a body of work where an influential philosophical theory of privacy
has informed the design of a logic of privacy. This logic has been used to 
formally specify a number of privacy regulations. Associated formal 
monitoring methods enable automated enforcement of parts of these regulations.
These results can be viewed as a more expressive counterpart in privacy to work on 
enforceable security policies~\cite{Schneider00}. 

Contextual integrity is a philosophical theory of privacy~\cite{Nissenbaum}. 
The building blocks of this theory are social contexts and context-relative informational norms. 
A context captures the idea that people act and transact in society not simply as individuals in an undifferentiated social world, but as
individuals in certain roles in distinctive social contexts, such as healthcare, education, friendship, and employment.
Norms prescribe and proscribe the flow of personal information in a given context, e.g., in a healthcare context a norm might prescribe flow of personal health information from a patient to a doctor and proscribe flows from the doctor to other parties who are not involved in providing treatment.
This theory has been used to explain why a number of technology-based systems and practices threaten privacy by violating entrenched informational norms. The theory is now well known in the privacy community and has influenced privacy policy in the US (for example, ``respect for context'' was included as an important principle in the Consumer Privacy Bill of Rights released by the White House in 2012~\cite{privacybillofrights}).

The idea that privacy expectations can be stated using context-relative informational norms is formalized in a 
semantic model and logic of privacy~\cite{BarthDMN06} and developed further in follow-up work~\cite{DeYoungGJKD10,GargJD11}. 
While contextual integrity talks about information flow norms in the abstract, a precise logic enables specification 
in a form that information processing systems can check for violations of such norms.  
Two considerations are particularly important in designing the logic: (a) expressivity --- 
the logic should be able to represent practical privacy policies; and (b) enforceability --- 
it should be possible to provide automated support for checking whether traces satisfy policies 
expressed in the logic. While the initial work of Barth et al.~\cite{BarthDMN06} employed 
first-order linear temporal logic (LTL) for specification, enforcement was limited to propositional 
LTL. Garg et al.~\cite{GargJD11} present an expressive enforceable logic of privacy. 
This privacy logic is an expressive fragment of first-order logic. It has been used to develop 
the first complete formalization of all disclosure-related clauses of two US privacy laws: the HIPAA Privacy
Rule for healthcare organizations and the Gramm-Leach-Bliley Act for financial institutions~\cite{DeYoungGJKD10}.
These comprehensive case studies shed light on common concepts that arise in information flow norms in 
practice---data attributes, dynamic roles, notice and consent (formalized as bounded time temporal properties), 
purposes of uses and disclosures, and principals' beliefs---as well as how individual norms are 
composed in privacy policies.  A related early effort on formal techniques to specify and analyze legal 
privacy policies appears in May et al.~\cite{MayGL06}.

At a technical level, the policy enforcement algorithm of Garg et al.~\cite{GargJD11} advances run-time monitoring 
formal methods to a restricted fragment of first-order logic. Chowdhury et al.~\cite{ChowdhuryJGD14} further 
improve the time- and space-efficiency of this algorithm by using a fragment of Metric First-Order Temporal 
Logic as the specification logic and using summary structures to compactly represent relevant state from 
the execution trace. Related formal methods are also employed in the work of Basin et al.~\cite{BasinKMZ15}.

\paragraph*{Privacy in statistical databases.} 

{\em Differential privacy}~\cite{dwork-sensitivity06, dwork-privacy06,
  dwork2014algorithmic} has emerged in the past decade as a gold standard
definition for strong privacy in statistical databases, giving rise to a veritable mountain of work
in both algorithms and systems conferences as well as many variations and
refinements.  The basic idea is that, by adding a small amount of random
noise to the result of an aggregate query over a large data set (e.g.,
``What fraction of the patients in this study were smokers but did not
develop cancer?''), we can guarantee that the presence or absence of any
single individual in the data set can make only a small difference in the
distribution of outputs---i.e., the privacy loss for any individual from any
differentially private query is bounded in a precise sense.  One major
attraction of differential privacy is that it is {\em compositional}: the
privacy loss from publishing the results of two differentially private
queries is no more than the sum of the losses for running either of the two
queries separately.  This avoids vulnerabilities of earlier privacy
definitions such as {\em k-anonymity}~\cite{k-anon}, where the results
of two separate privacy-preserving queries can be combined to completely
violate privacy, as happened in the Netflix Challenge
debacle~\cite{narayanan-2008}.

There are now a number of {\em query languages for differentially private
  data analysis}, including Pinq~\cite{pinq, proserpio2012workflow, Ebadi:2015:DPG:2676726.2677005},
Airavat~\cite{roy-2010-airavat}, DJoin~\cite{narayan-2012-osdi},
Fuzz~\cite{fuzz, haeberlen-2011-usenixsec}, DFuzz~\cite{gaboardi2013linear},
VFuzz~\cite{narayan2015verifiable}, GUPT~\cite{mohan2012gupt}, and others.
The goal of all these languages is to automatically enforce privacy
restrictions, allowing the owners of sensitive datasets to query them (or
make them available for querying by others) without fearing that mistakes or
malicious intent will lead to privacy breaches.

The languages mentioned above make it easy to query sensitive data
without fear of violating privacy, but they are also limited in that each
embodies a specific ``format'' for private queries.  By contrast, the
algorithms literature is full of complex and subtle methods for answering
particular sorts of questions while guaranteeing differential privacy, and
the majority of these algorithms fall outside the scope of what can be
expressed and automatically verified by these languages.  This has led to
another thread of work that has demonstrated promising initial successes in
going beyond fully automatic enforcement and into the realm of {\em
  interactive verification tools for privacy-preserving computations}.
Gilles Barthe's group at IMDEA is probably furthest ahead in this area; for
example, their CertiPriv system~\cite{Barthe:2013:PRR:2542180.2492061} has
been used to verify a number of examples whose formal analysis is out of the
reach of previous techniques. In particular, they give the first
machine-checked correctness proofs for the Laplace, Gaussian, and
exponential mechanisms (three critical building blocks for differentially
private algorithms) and of the privacy of some recent randomized and
streaming algorithms.

\paragraph*{Deployed privacy-preserving systems.} 

Besides these academic successes, some significant success stories are
starting to come from industrially deployed privacy-protection systems that are
either directly supported by formal methods or simply inspired by more
formal work.

Recent work~\cite{SenGDRTW14}  develops a formal methodology and tool chain for 
checking software systems written in big data programming languages 
(e.g., Scope, Hive, Dremel) for compliance with a class of privacy policies. 
The privacy policies restrict direct and implicit information flows based on 
role, purpose, and other considerations. The tool chain has been applied to 
check over a million lines of source code in Microsoft Bing's data analytics 
pipeline for compliance with its privacy policies. This work addresses two 
central challenges in making privacy compliance tools practical. First, it 
presents the \emph{Legalease} policy language that allows precise 
specification of real-world privacy policies while still being usable by 
the target users of this language---the legal privacy team. Second, it 
presents the \emph{Grok} data inventory tool that maps existing code-level
schema elements to datatypes in Legalease, in essence annotating existing
programs with information-flow types with minimal human input. Compliance
checking is then reduced to a form of information-flow analysis of big 
data programs. The design of Legalease (especially its treatment of nested 
allow-deny rules) was influenced, in part, by prior work on logical formalization of 
the HIPAA Privacy Rule~\cite{DeYoungGJKD10} mentioned earlier in this section. 
The compliance checking method 
was influenced, in part, by work on language-based privacy~\cite{HayatiA04}. 

Several types of privacy-preserving systems have recently been deployed 
at scale. While not supported yet by formal methods, we mention them 
here because they serve as useful motivation for basic research in this area. 
They are also attractive targets for application of the already developed 
formal methods. 

For example, the formal methods developed to support 
differentially private data release could be directed to the study of 
the design and implementation of the RAPPOR system from Google~\cite{ErlingssonPK14}. 
Another example is the system for differentially private release of 
password frequencies that was recently employed to release statistics 
about 70 million user passwords by Yahoo!~\cite{BlockiDB16}.

While much of our focus here has been on data privacy, another significant
area of privacy research is communication privacy, where a significant body
of work has emerged on anti-sur\-veil\-lance tools and their foundations. An
influential and widely used tool in this area is
Tor~\cite{DingledineMS04}. While there is some work on formal analysis of
anonymous communication protocols and systems including
Tor~\cite{Shmatikov04,FeigenbaumJS12}, this is a rich area that awaits a
deeper dive from our community.

\subsection{Grand Challenges}
\label{grandchal}

These successes offer encouraging evidence that formal methods may
fruitfully be applied to technologies for preserving privacy.  But a great
deal remains to be done, both at the level of conceptual and mathematical
foundations and at the level of deployable technologies.  In
this section, we identify some of the main foundational challenges,
and outline several potential grand
challenge applications that could drive further foundational advances at
the same time as learning how to deal with engineering and organizational
hurdles and delivering useful systems.

\paragraph*{Grand challenge 1:} Map the privacy space. This
foundational 
challenge demands a deeper understanding of the concept of privacy and
its relationship to neighboring concepts in the personal data
protection space, such as fairness and transparency. This viewpoint is
consistent with conceptions embodied in the Fair Information Practices
Principles (FIPPs)~\cite{oecd1980} and in recent reports from the
White House~\cite{whitehouse2014bigdata}. Often, privacy and these
related properties can be understood as imposing different kinds of
information flow and use constraints. The grand challenge involves
developing computational formalizations of a broad range of these
properties and studying their relationships, much as we have a broad
set of security definitions in cryptography and a formal understanding
of their relationships.

\paragraph*{Grand challenge 2:} Develop and deploy {\bf privacy-preserving tools for scientific discovery}, that is, data exploration and analysis tools that can be used by medical researchers, social scientists, and other academics working in data-intensive fields to carry out their daily work.  Until now, most research in social and life sciences takes a fairly rough-and-ready approach to privacy, while work on strong notions of privacy (e.g., differential privacy) and accompanying tools has not made much impact outside of computer science.  The goal would be the publication of papers in strong subject-area journals whose results are obtained by analyzing real data sets using a research analytics system with strong, formally verified privacy guarantees. 

\paragraph*{Grand challenge 3:} Develop foundations and tools that support {\bf privacy and accountability in big-data analytics}.  Contemporary research in, for example, health-care often proceeds by attempting to learn models from large, privacy-sensitive datasets.  This methodology raises two competing concerns.  First, public release of these models themselves (for example, in academic publications) may violate the privacy of individuals whose data is included in the studies.  Second, to support future research or clinical practice, these models must be transparent, or explainable---i.e., it must be evident what features of the data led to particular conclusions.  The challenge here is to improve transparency of big-data analytics (a difficult problem in itself!) while still preserving privacy.  Socially relevant applications abound: online personalization, predictive policing, credit scoring, insurance risk estimation, etc.  The goal would be to influence the design and analysis of industrial systems in these areas. More generally, accountability in big-data analytics demands methods for detecting violations of privacy, explaining how these violations came about, assigning responsibility and blame, and then adopting appropriate corrective measures. The call for accountability in big data analytics and its importance for protecting privacy and other values is being increasingly recognized~\cite{whitehouse2014bigdata,Diakopoulos16}, with initial results beginning to appear in the privacy literature~\cite{datta2015opacity,LecuyerSSCGH15,datta2016transparency}.

\paragraph*{Grand challenge 4:} Develop methods for {\bf balancing privacy and accountability in protocols}.  A clear case of the need for this is in voting systems: we want to develop protocols that protect sensitive information such as who voted for which candidate while making it possible for election officials to audit election results.  Similarly, currencies for crypto-currencies need to maintain anonymity while ensuring that, if someone tries to spend the same ``coin'' twice, either they will not succeed or they will be detected.  Accountability is enforced either through post-hoc blame assignment or through economic incentives that deter misbehavior.  One concrete goal would be a formal, comprehensive privacy and accountability analysis of a widely used crypto-currency such as BitCoin.  Another goal---already actively pursued, but worth reiterating---is a formal analysis of a deployed voting system.  Another is anonymous communication, where it is desirable to be able to tie actions to individuals under some cases (such as misbehaving users or illegal activity) while preserving anonymity under ordinary circumstances.  Another is formal analysis of anonymous credentials, which can be used to prove that some property of an individual (being more than 21 years old, belonging to some organization, having paid to drive on a particular set of toll roads) while not revealing identity.  Mechanized verification of such protocols is a crosscutting challenge.

\paragraph*{Grand challenge 5:} Develop fundamental concepts and formally
verified, deployable technologies for protecting privacy in {\bf
  cyber-physical systems} such as the {\bf Internet of Things}.  This domain raises some particularly 
  significant challenges for formal methods. First, since we envision a world in which user data will be collected 
  and used by numerous devices, it will be particularly important to be cognizant of user preferences during 
  privacy enforcement. This observation necessitates developing usable languages for expressing privacy preferences 
  (by users) and infrastructure-side privacy requirements (by developers), formally connected to enforcement mechanisms.  
  Second, these systems, while controlled by software, will interact continuously with the physical world. Thus, 
  privacy models have to be aware of the interaction between software systems and physical dynamical systems and 
  enforcement methods and their formal verification will require us to go significantly beyond the state-of-the-art 
  in the cyberphysical systems (CPS) area. A concrete challenge is that while much prior work in CPS has focused on 
  safety verification, privacy verification will require advances that go beyond reasoning about trace properties.

\section{Cross-Cutting Considerations}

\subsection{Whole-system guarantees}
\label{ss:whole-system}

Current formal method techniques can provide strong security
guarantees for individual components of a system, and at varying
levels of abstraction. Ideally, however, we
want security guarantees for the whole system.

Systems are built by composing existing libraries, sub-systems, and
components. For example, a system that provides a
    web service may comprise a web server, a database, a web
    application, and the Linux operating system. Each of these
    sub-systems are themselves composed of many components.

Security issues can arise at the boundaries between components, even
though individual components may be ``secure'' (e.g., \cite{Mantel2002,DattaFGJK2011}). 
Security
issues at these boundaries can be exacerbated when
different individuals or organizations have responsibility for the
various
components. 

In addition to ensuring security guarantees when individually-secure
components are composed together,  ``whole-system guarantees'' (also ``cross-layer security'' and sometimes ``end-to-end guarantees'') refers
to security guarantees that hold across abstraction boundaries. For
example, having assurance that a system is secure at all levels of
abstraction, from the hardware through the operating system, through
the network/distribution layer, to the application itself. This
requires, for example, that assumptions made at the application-level
are in fact guaranteed to be enforced by lower-level abstractions.

There have been several promising success stories (and in-progress stories) for whole-system
security, including Ironclad~\cite{hawblitzel:ironclad}, the HACMS DARPA program~\cite{hacms-catalog}, and
verification of a radiation therapy system~\cite{ErnstGJLPTTW2015}.

\medskip

However, significant challenges must be overcome to use formal
methods to provide whole-system security guarantees.

Incompatibility of formal method tools can hamper 
integration of individually-secure components. Currently there are
a few standards for low-level tools such as SAT or SMT solvers~\cite{BarFT-RR-15,Starexec}, but
no
common standards for formal specifications and proofs, and existing
tools can vary greatly in their representation of
specifications and proofs. Thus if different tools are used to
formally validate the security of components, it may require significant effort
to combine these formalizations. Similarly, there is a lack of
standardization of threat models and formal security guarantees,
and mismatches in the
statement of security guarantees that individual components achieve
can complicate achievement of whole-system security guarantees. 

One instance of such a mismatch is between standard
cryptographic-style security specifications 
and traditional program logics. 
Cryptographic
security specifications and proofs typically require pervasive
reasoning about probabilities. This is typically at odds with the compositional structure of program logics. A notable exception is universal composability~\cite{Canetti2001}, %
a framework for cryptographic protocols that preserves security under composition. However, universal composability is a very strong requirement that is difficult to achieve in cryptographic protocols.

\medskip

Difficulties with providing whole-system security guarantees across
abstraction layers may be indicative that current abstractions do not
and can not provide security guarantees that are sufficient to satisfy
security assumptions required at higher-levels of
abstraction. For example, TCP does not provide
  any liveness guarantees, making it difficult or impossible to
  provide whole-system liveness properties in systems that use
  TCP.

As we develop our understanding of the formal guarantees that
abstraction layers require and are able to provide, we may identify
opportunities to modify the abstraction layers to improve the use of
formal methods for security and privacy. That is, clean-slate
approaches to the design of whole systems may enable whole-system
security using modular formal methods.

\paragraph{Challenge problems} Simple systems may provide suitable
challenge problems to both highlight difficulties of providing
whole-system security guarantees, and also to advance the
state-of-the-art of (modular and composable) formal methods for whole-system security.
We propose two such systems as challenge problems.

  \begin{itemize}
  \item \emph{Develop a formally verified crypto-currency wallet.} Crypto-currencies such as BitCoin and Eth\-er\-eum use distributed cryptographic mechanisms to secure financial transactions and the creation of new units of currency. Users of a crypto-currency rely on software---called a \emph{wallet}---to store, send, and receive currency. The financial relevance of wallets provides a clear motivation to provide strong whole-system security guarantees for wallets (including characterization and enforcement of privacy and accountability; see Section~\ref{grandchal}). The emergence of \emph{hardware wallets} (that use dedicated or specialized hardware) means that it may be feasible to provide strong security guarantees from the hardware upwards. Wallet software interacts with distributed cryptographic mechanisms in interesting ways, which adds an additional dimension to this challenge problem.

  \item \emph{Develop an end-to-end secure messaging system on a peer-to-peer overlay.} Messaging systems---such as text messages over a phone network, or instant messaging systems---are increasingly used to send sensitive information. Adversaries may seek to learn confidential messages, corrupt messages, or learn the senders and recipients of messages. The security of a messaging system depends on the protocols of the distributed system and also on the design and implementation of end-user software. 
    
\end{itemize}

\subsection{Education and outreach}\label{ss:education}

Use of formal methods for security and privacy can be
encouraged by increased awareness of the need for strong security and privacy
guarantees and of the availability and capabilities of formal
methods. While there is significant potential for outreach to software
practitioners, graduate students, and K--12 students, we focus here on
undergraduate students.

Many colleges offer courses in computer security, and some even offer
degrees specializing in computer security (e.g., University of
Maryland, University of Delaware, and Boston University). Similarly,
there are a number of courses devoted to formal methods (see the report from the 2012
NSF Workshop on Future Directions for Formal Methods and Its
Transition To Practice~\cite{formalmethodsreport} for a list of
college courses that focus on formal methods). However, in general,
these courses are electives and taken by a relatively small proportion
of undergraduate students. To increase the use of formal methods for
security in the next generation of software practitioners, it is
desirable for most students completing a computer science
undergraduate degree to be exposed to both the need for strong security
and privacy
guarantees, and the capabilities of formal methods.

The Computer Science Curricula 2013 \cite{CS2013}---developed by
the ACM/IEEE-CS Joint Task Force on Computing Curricula---added
\textit{Information Assurance and Security} as a new knowledge area
and notes that it ``is unique among the set of [Knowledge Area]s'' in
that its ``topics are pervasive throughout other Knowledge Areas.''
This reflects the cross-cutting nature of security. To
the extent that the Computer Science Curricula are incorporated into
college-level classes, we hope to see an increase in the students that
encounter core security and privacy concepts during their undergraduate
education.

Formal methods are also cross-cutting, providing tools and techniques
that support the correct design and implementation of many kinds of
systems. In the Computer Science Curricula 2013, however, formal methods are
presented only as an elective topic (in the Knowledge Area of Software
Engineering). 
Introduction of courses specifically focused on formal
methods is likely not the best approach to expose a broad range of
students to the concepts and benefits of formal methods. Such courses
would likely be electives and taken by a relatively small number of students. Instead,
incorporation of formal methods and tools into existing courses can
show how formal methods can support better design and implementation,
for example, proving properties of distributed systems, or verifying
the correctness of hardware design. This approach has been used
successfully in several courses, such as at Northeastern University,
where second semester freshmen use ACL2 to
  state and prove properties about the programs they
  write~\cite{EastlundVF2007}. This approach is also the
  recommendation of the 2012 NSF Workshop on Formal Methods~\cite{formalmethodsreport}.

  We do not expect most colleges and universities to have sufficient
  resources to offer specialized courses in formal methods applied to
  security and privacy. Rather, security courses offer an opportunity
  to present formal methods in context, supporting the design and
  implementation of secure systems. This has been the approach taken
  in, for example, Professor Michael Hicks' (University of Maryland,
  College Park) Coursera MOOC on Software
  Security\footnote{\url{https://www.coursera.org/course/softwaresec}},
  which has units on the use of static analysis and symbolic execution to
  build secure software.

\subsection{Tools, usability, infrastructure}\label{ss:tools}

Tools are vastly more capable than a few years ago. Both
satisfiability solvers (SAT solvers; software that, given a
propositional formula, will identify an assignment of truth values
that satisfies it) and satisfiability-modulo-theory solvers (SMT
solvers; satisfiability solvers that find an assignment of values
subject also to theories such as linear arithmetic) are vastly stronger.
Theorem provers are rigorous, and a body of experience shows
that they can be successfully used on a large scale.  Code analysis
techniques---such as ensuring that code respects given invariants,
including type systems---are better, and well-integrated into solvers
or provers.  Theorem provers and SMT solvers may be used in
combination, providing human assistance where SMT solvers are weak,
such as reasoning in non-linear arithmetic, as needed in cryptography.

Nevertheless, workshop participants are painfully aware of the quirks
and limitations of tools that support formal methods.  Renaming
variables may sometimes cause a SAT solver to fail; changes to
specifications that reorder premises can cause proof scripts to fail.
This creates a challenge in incorporating formal methods into system
build processes, for which stability is important.  Worse,
verification toolchains transform the user's input substantially, with
the consequence that errors are often reported in terms remote from the
user's source code.

The participants explored various paradigms for the relation between
formal artifacts and testing.  Many organizations (though not all)
maintain systematic unit tests to control software evolution.  Unit
tests interact with specifications in desirable ways.  When unit tests
already exist, they can be used to check whether the specifications
express the intended goals.  Alternatively, many unit tests can be
generated from specifications; this helps to provide empirical
evidence that code meets its specifications.  Especially, it provides
empirical evidence when it doesn't, and this evidence can be much
quicker to produce than rigorous analysis.  This type of counterexample
may be much harder to construct formally, although finite model finders (e.g., Alloy~\cite{jackson12:alloy}) and
bounded model checkers~\cite{clarke1999model} are geared toward this goal.

\section{Conclusion}

Cyber attacks threaten personal privacy, economic activity, society's
infrastructure, and national security.  The game is asymmetric:  An
attacker has a wide choice of strategies, which may use a succession
of footholds traversing different abstraction layers.  Because
attribution is difficult, many exploratory sallies may precede a
successful attack.  Thus, the defender must win against every
strategy, while the attacker need find only a single one.  

This is the core reason formal methods are indispensable to
security.  Formal methods reason about computational entities, using
logical or mathematical descriptions of the entities to draw reliable
conclusions about the behavior of those entities.  Without this
modeling rigor, systems of realistic complexity will always offer
behaviors an adversary can exploit.  Moreover, after a long period of
development, formal methods now provide strong techniques for a
variety of different styles of modeling.

In this report, we have advocated their systematic role in security
research, with increasing impact on the development of our secure
software and hardware.  Widened use of formal methods will provide techniques
calibrated to specific security goals, and establish that specific
types of systems meet those goals.  These efforts will also make specific,
formally certified components available, which can be incorporated
into new systems.  This process will ease the burden of future formal
secure development, leading to an acceleration of productivity and
wide increases in the quality of secure systems.  Methods that lack
rigor will never lead to comparable improvements, since they provide
no overview of the attacker's possible strategies, and no evidence to
exclude their success.

Formal methods include a wide range of techniques, tools, and
approaches, and these should be flexibly applied.  Not all systems aim
at the same security goals, and researchers should be explicit about
the properties that they intend to achieve.  The nature of these
properties helps to determine what formal methods are appropriate, and
what balance to strike among specification, rigorous hand proof, and
mechanized proof support.

We should not underestimate the challenges.  Many security problems
arise from the interactions of different layers in a system's stack,
leading from hardware through kernel and networking infrastructure
toward applications.  These layers often use quite different
abstractions, and using the specification of the services of one layer
to discharge the assumptions of the next higher layer often involves
guesswork.  Indeed, identifying the goals of different stakeholders at
a given layer often involves psychoanalyzing their use cases.
Extensive experience and, most likely, redesign of components well
entrenched in today's systems will be needed.

We have identified pressing research topics---such as \emph{whole-system guarantees}, choice of \emph{abstractions}, finding
\emph{compatible} tools, proofs, and modeling styles, and
\emph{development methods} usable in practice---and made them concrete
by proposing \emph{grand challenge} efforts that will stimulate
resolving them under the constraints of important, realistic outcomes.
Finally, we emphasize that improved education will be needed, with
better training in security and formal methods both; and that the
community will need to build up reusable infrastructure and tools.
The result will be transformational improvements in the security of
systems on which our society relies.

\section*{Acknowledgments}

We thank the National Science Foundation for sponsoring the workshop
and Program Managers Nina Amla and Anindya Banerjee for advice and
discussions.
We thank Michael Hicks for hosting the workshop at UMD College Park,
and Tina Knight for administrative and logistical support. We thank
Owen Arden, Tej Chajed, Justin Hsu, and Joseph McMahan for scribing the
workshop discussions.
We sincerely thank the workshop participants. Their active and
enthusiastic engagement produced the core results of this report.
The participants were:
Nina Amla,
Owen Arden,
Anindya Banerjee,
Tej Chajed,
Steve Chong,
John Criswell,
Anupam Datta,
Steven Drager,
Nate Foster,
Matt Fredrikson,
Sol Greenspan,
Carl Gunter,
Aarti Gupta,
Joshua Guttman,
Michael Hicks,
Justin Hsu,
Bart Jacobs,
Somesh Jha,
Frans Kaashoek,
Daniel Kifer,
Boris K\"{o}pf,
Rao Kosaraju,
Shriram Krishnamurthi,
Petros Maniatis,
Z. Morley Mao,
David Mazi\`{e}res,
Jon McCune,
Patrick McDaniel,
Bill McKeever,
Joseph McMahan,
Ken McMillan,
Toby Murray,
Andrew Myers,
Bryan Parno,
Benjamin Pierce,
Phil Regalia,
Patrick Schaumont,
Deborah Shands,
Zhong Shao,
Tim Sherwood,
Elaine Shi,
Cynthia Sturton,
Jakub Szefer,
Cesare Tinelli,
David Walker,
Xi Wang, and
Nickolai Zeldovich.

\bibliographystyle{plainnat}
\bibliography{bib}

\end{document}